\documentclass[aps,prb,groupedaddress,twocolumn, floatfix,showpacs]{revtex4}
\usepackage{graphicx,amsmath,amssymb,bm,hyperref,color}
\definecolor{blue}{rgb}{0.0,0.0,1.0}
\definecolor{black}{rgb}{0.0,0.0,0.0}
\definecolor{red}{rgb}{1.0,0.0,0.0}
\hypersetup{colorlinks=true,breaklinks, linkcolor=blue,urlcolor=blue,citecolor=red}
\begin{document}

\title{On the shot-noise limit of a thermal current}
\author{Kelly R. Patton}
\email[\hspace{-1.4mm}]{kpatton@physnet.uni-hamburg.de}
\affiliation{I. Institut f\"ur Theoretische Physik Universit\"at Hamburg, Hamburg 20355, Germany }
\date{\today}
\begin{abstract}
The noise power spectral density of a thermal current between two macroscopic dielectric bodies held at different temperatures and connected only at a quantum point contact is calculated. Assuming the thermal energy is carried only by phonons, we model the quantum point contact as a mechanical link, having a harmonic spring potential. In the weak coupling, or weak-link limit, we find the thermal current analog of the well-known electronic shot-noise expression.
\end{abstract}
%\keywords{}
\maketitle
\section{Introduction}
Just as Ohm's law, relating the electrical current to an applied potential, breaks down when the quantum mechanical aspects of the charge carriers becomes important, such as in the mesoscopic regime; Fourier's Law of heat conduction suffers a similar fate.  Mesoscopic phonon systems\cite{Clelandbook} provide some of the best experimental setups to test the quantum nature of heat transport,  such as the quantization of thermal conductance.\cite{SchwabNature00}
Although, experimental demonstration of which lagged a decade behind that of its electronic counterpart.\cite{vanWeesPRL88}

 Following this seminal work, nanomechanical systems have since seen an increased interest, experimentally and theoretically, from such diverse areas as  quantum computing\cite{ClelandPRL04} to promising research into detecting the quantum mechanical zero-point motion of a macroscopic object.\cite{BlencoweScience04, LaHayeScience04} Similar to the quantization of electrical conductance, where each channel of a one-dimensional conductor can contribute a quantum of electrical conductance, ${e^{2}}/{2\pi \hbar}$ per spin, in a one-dimensional dielectric each vibrational mode carries a quantum of thermal conductance given by $\pi k^{2}_{\rm B}T/6\hbar$.  Of course one requirement to observed this quantization is a clean system with minimal scattering, i.e.~ballistic transport.  Within the Landauer-B\"uttiker formalism, this amounts to setting the transmission matrix to unity for each mode.  The opposite limit of weak transmission or strong scattering can be equally interesting.  For instance in a system of two conductors separated by a thin tunnel barrier, such as a scanning tunneling microscope (STM), the electrical conductance, associated with the tunneling current, is related to the product of the local density of states on each side of the barrier.\cite{Mahanbook} In Ref.~[\onlinecite{PattonRBR01}] a thermal analog of an STM, i.e.~a phonon mediated scanning {\it thermal} microscope, was proposed, where the thermal conductance associated with the energy current between two macroscopic dielectric bodies held at different temperatures and connected at a single quantum point contact was found to be related to the the local phonon density of states of each reservoir.  Similar work has been done involving the the phonon dominated thermal transport through more complex connections, such as molecules.\cite{BuldumEurophysLett99, ChenJHeat02,SegalJChemPhys03} 
 
 Here we examine the noise of a thermal current in this limit of weak transmittance, the shot-noise limit.  In the same way the granularity of the charge carriers, in say a weak tunneling current, contributes to the current noise, the analogous behavior for phonons should be observed in a thermal current.
\footnote{Due to the bosonic nature of phonons, distinguishing the total energy carried by a single phonon or two or more with smaller energy would be difficult to discern.} Experimentally, the ability to detect a single phonon is an ongoing  area of interest. \cite{RoukesPhysicaB99}

 In Ref.~[\onlinecite{PattonRBR01}] the thermal current between two insulators weakly joined by only a mesoscopic link, modeled as a harmonic spring, was calculated.  The actual physical link  could be a few chemical bonds or even a small bridge of material, see Fig.~[\ref{fig1}].  The result of Ref.~[\onlinecite{PattonRBR01}] was the thermal analog of the well-known tunneling current formula.\cite{Mahanbook} In the following sections we examine the intrinsic noise present in such a thermal current.  It is assumed the two bodies are only weakly coupled, to lowest order in the coupling, this is equivalent to the shot-noise limit of the electronic counterpart.  
 
\section{Model and Thermal Current} 
\begin{figure}
\includegraphics[scale=.50]{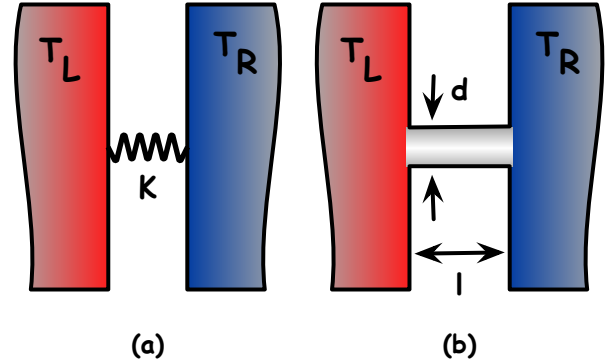}
\caption{(color online) Model of two macroscopic  dielectric bodies held at different temperatures $T_{\rm L}$ and $T_{\rm R}$ and joined by a single quantum point contact. The contact can be taken to be (a) several atomic bonds with a spring constant $K$ or (b) a small ``neck'' of dialectic material of length $l$ and diameter $d$ with an effective spring constant $K=(\pi d^{2}/4l)Y$, where $Y$ is the Young's modulus of the material.  \label{fig1}}
\end{figure}
We consider the following model, which is illustrated in Fig.~[\ref{fig1}]:  Two macroscopic dielectric bodies, labeled left (L) and right (R), act as thermal reservoirs and are held at fixed temperatures $T^{}_{\rm L}$ and $T^{}_{\rm R}$.  The Hamiltonian for each side is taken, in the harmonic approximation, as $(\hbar=1)$
\begin{equation}
H^{}_{I}:=\sum_{n}\omega^{}_{In}a^{\dagger}_{In}a^{}_{In},\hspace{.5cm} I={\rm L,R}
\end{equation}
where $a^{\dagger}_{In}$ and $a^{}_{In}$ are phonon creation and annihilation operators for the left and right side, which satisfy 
\begin{equation}
\big[a^{}_{In},a^{\dagger}_{I'n'}\big]=\delta_{nn'}\delta_{II'}
\end{equation}
and 
\begin{equation}
\big[a^{}_{In},a^{}_{I'n'}\big]=\big[a^{\dagger}_{In},a^{\dagger}_{I'n'}\big]=0.
\end{equation}
The quantum point contact, or weak mechanical link,  is modeled as a harmonic potential with spring constant $K$
\begin{equation}
\delta H:=\frac{1}{2}K\big[u^{z}_{\rm L}({\bf r}_{0})-u^{z}_{\rm R}({\bf r}_{0})\big]^{2},
\end{equation}
where ${\bf r}_{0}$ is the point of contact bewteen the two reservoirs and $u^{z}_{I}$ is the $z$-component (the direction normal to each surface) of displacement field ${\bf u}_{I}({\bf r})$.  This model of the weak link assumes the compressional strength of the link dominates over other such displacements such as flexorial or torsional.  In principle these interactions could also be included, this would amount to replacing the spring constant $K$ with a tensor quantity coupling to different components of the phonon field operator.

The displacement field of each reservoir can be expanded in terms of phonon creation and annihilation operators as
\begin{equation}
 {\bf u}_{I}({\bf r}):=\sum_{n}\sqrt{\frac{1}{2\rho\omega_{In}}}\big[a^{}_{In}{\bf f}^{}_{In}({\bf r})+a^{\dagger}_{In}{\bf f}^{*}_{In}({\bf r})\big],
 \end{equation}
where ${\bf f}_{In}({\bf r})$ are the normalized vibrational eigenfunctions, and $\rho$ is the mass density. 
\subsection{Thermal Current}
\label{thermal current section}
Because of energy conservation and using Heisenberg's equation-of-motion, a thermal-current operator can be defined as
\begin{equation}
{\hat I}_{\rm th}:=\partial_{t}H_{\rm R}=i\big[H,H_{\rm R}\big], 
\end{equation}
where the full Hamiltonian $H=H_{\rm L}+H_{\rm R}+\delta H$. 
Performing the commutator gives  
\begin{equation}
\label{thermal current operator}
\hat{I}_{\rm th}=\frac{iK}{2}\sum_{nn'}\omega_{{\rm R}n}\big\{A_{{\rm R}n'}-A_{{\rm L}n'},h^{}_{{\rm R}n}a^{}_{{\rm R}n}-h^{*}_{{\rm R}n}a^{\dagger}_{{\rm R}n}\big\},
\end{equation}
where $h_{In}:=({2\rho\omega_{In}})^{-1/2}f^{z}_{In} $, $A_{In}:=h_{In}a_{In}+h^{*}_{In}a^{\dagger}_{In}$, and $\{\cdot,\cdot\}$ is the anticommutator. 
Treating the coupling as the perturbation; within linear response, the thermal current is\cite{PattonRBR01} 
\begin{equation}
\label{thermal current}
\big<\hat{I}_{\rm th}\big>=2\pi K^{2}\int\limits_{0}^{\infty}d\epsilon\,\epsilon\, N^{zz}_{\rm L}({\bf r}_{0},\epsilon)N^{zz}_{\rm R}({\bf r}_{0},\epsilon)\big[n^{\rm B}_{\rm L}(\epsilon)-n^{\rm B}_{\rm R}(\epsilon)\big],
\end{equation}
where $n^{\rm B}_{I}(\epsilon)=(\exp(\epsilon/k_{\rm B}T^{}_{I})-1)^{-1}$ is the Bose distribution and 
\begin{equation}
\label{spectral density}
N^{zz}_{I}({\bf r},\omega)=\sum_{n}|h_{In}({\bf r})|^{2}\delta(\omega-\omega_{nI})
\end{equation}
is the $zz$ component of the local {\it spectral} density.  It should be noted that Eq.~(\ref{spectral density}) is {\it not} equal to the $zz$ component of the local  phonon density of states tensor given by,\cite{GellerPRB04} 
\begin{equation}
g^{ij}_{I}({\bf r},\omega)=\sum_{n}f^{i}_{In}({\bf r})\big[f^{j}_{In}({\bf r})\big]^{*}\delta(\omega-\omega_{nI}),
\end{equation}
but $N^{zz}_{I}({\bf r},\omega)$ is related to the imaginary part of the retarded phonon Green's function and is the relevant quantity of interest for the present work. For clarity the superscripts $zz$ will be dropped from here on.   Eq.~(\ref{thermal current}) is the thermal analog of the expression for an electronic tunneling current, Eq.~(\ref{electronic tunneling current}).

\section{Calculation of the Phonon Shot-Noise}
Here we calculate the intrinsic noise\footnote{The noise generated by the system of study and not from external experimental equipment.}  associated with a thermal current as calculated in Sec.~\ref{thermal current section}. 
The symmetrized  noise is defined as\cite{noisebook1,noisebook2, BlanterPhysRep00} 
\begin{equation}
S_{\rm th}(\omega):=\frac{1}{2}\int dt\, e^{i\omega t}\big<\big\{\delta \hat{I}_{\rm th}(t),\delta \hat{I}_{\rm th}(0)\big\}\big>_{H},
\end{equation}
where $\delta  \hat{I}_{\rm th}:=\hat{I}_{\rm th}-\big<\hat{I}_{\rm th}\big>_{H}$.\footnote{Sometimes the factor of 1/2 is omitted and thus will change some subsequent formulas by a factor of two.} In Ref.~[\onlinecite{BlencowePRB99}] the short time, or high-frequency ($\omega\rightarrow\infty$), noise of a general heat current was studied.  Here we investigate the long time or low-frequency ($\omega\rightarrow0$) noise in the weak transmission limit.  

 To lowest order in the interaction the noise is simply
\begin{equation}
\label{general shot noise}
S_{\rm th}(\omega):=\frac{1}{2}\int dt\, e^{i\omega t}\big<\big\{\hat{I}_{\rm th}(t),\hat{I}_{\rm th}(0)\big\}\big>_{H_{0}},
\end{equation}
where $H_{0}:=H_{\rm L}+H_{\rm R}$,
\begin{equation}
\big<\hat{O}\big>_{H_{0}}=\frac{{\rm Tr}\,e^{-\beta H_{0}}\hat{O}}{{\rm Tr}\,e^{-\beta H_{0}}},
\end{equation}
and
\begin{equation}
\hat{O}(t)=e^{iH_{0}t}\hat{O}e^{-iH_{0}t}.
\end{equation}
Using Eq.~(\ref{thermal current operator}), along with dropping anomalous terms, the zero-frequency component of the noise is\footnote{Because the phonons are noninteracting, in the harmonic approximation used here, the  correlation functions involved can easily be evaluated.} 
\begin{widetext}
\begin{equation}
\label{noisea}
S_{\rm th}(\omega=0)=2\pi K^{2}\int\limits_{0}^{\infty}d\epsilon\, \epsilon^{2}N^{}_{{\rm L}}(\epsilon) N^{}_{{\rm R}}(\epsilon)\Big\{n^{\rm B}_{\rm L}(\epsilon)\big[1+n^{\rm B}_{\rm R}(\epsilon)\big]+n^{\rm B}_{\rm R}(\epsilon)\big[1+n^{\rm B}_{\rm L}(\epsilon)\big]\Big\}
\end{equation}
or
\begin{equation}
\label{noiseb}
S_{\rm th}(\omega=0)=2\pi K^{2}\int\limits_{0}^{\infty}d\epsilon\, \epsilon^{2}\coth\Big[\frac{\epsilon}{2k_{\rm B}}\Big(\frac{1}{T_{\rm R}}-\frac{1}{T_{\rm L}}\Big)\Big]N^{}_{{\rm L}}(\epsilon) N^{}_{{\rm R}}(\epsilon)\big[n^{\rm B}_{\rm L}(\epsilon)-n^{\rm B}_{\rm R}(\epsilon)\big].
\end{equation}

It is illustrative to compare Eq.~(\ref{noiseb}) to the electronic expression of the zero-frequency component of the shot-noise,
\begin{equation}
\label{electronic shot noise}
S_{\rm el}(\omega=0)=e \big<\hat{I}_{\rm el}(eV)\big>\coth(eV\beta/2),
\end{equation}
where for a tunneling current
\begin{align}
\label{electronic tunneling current}
\big<\hat{I}_{\rm el}(eV)\big>=&2\pi e |T|^{2}\sum_{\sigma}\int d\omega\, \rho^{}_{\rm L}({\bf r}\sigma,\omega-eV)\rho^{}_{\rm R}({\bf r}\sigma,\omega)\big[n^{\rm F}_{\rm L}(\omega-eV)-n^{\rm F}_{\rm R}(\omega)\big]. 
\end{align}
\end{widetext}
Here $|T|^{2}$ is the transmission probability, $\rho_{I}(\omega)$ is the electronic local density of states, and $n^{\rm F}_{I}(\omega)$ is the Fermi distribution function. 

Assuming, as in most cases of interest, the phonon spectral density goes as a power-law at low energies, $N^{}_{I}(\omega)\propto\omega^{\alpha}$ and letting $T^{}_{\rm R}\rightarrow0$ for simplicity, the temperature dependance of the noise is given as
\begin{equation}
S_{\rm th}(\omega=0)\propto T^{3+2\alpha}.
\end{equation}
\subsection{Equilibrium noise }
In the limit $T^{}_{\rm L}\rightarrow T^{}_{\rm R}$ there is no net heat current; nonetheless, there remains  thermal fluctuations given by
\begin{equation}
\label{equilibrium noise}
S_{\rm th}(\omega=0)=2k_{\rm B}T^{2}G_{\rm th},
\end{equation}
where
\begin{equation}
G_{\rm th}:=\lim_{T_{\rm L}\to T_{\rm R}}\frac{I_{\rm th}}{T_{\rm L}-T_{\rm R}}=2\pi K^{2}\int\limits_{0}^{\infty}d\epsilon\, \epsilon\, N^{}_{{\rm L}}(\epsilon) N^{}_{{\rm R}}(\epsilon)\frac{\partial n^{\rm B}_{}(\epsilon)}{\partial T}
\end{equation}
is the linear thermal conductance.  This is the phonon analog of Nyquist-Johnson noise. In  an electronic system the Nyquist-Johnson noise is given by
\begin{equation}
S_{\rm el}(\omega=0)=2k_{\rm B}TG_{\rm el}.
\end{equation}
It should be noted that, in general Eq.~(\ref{equilibrium noise}) is a universal relation, regardless of the model used here, and is a consequence of the fluctuation-dissipation theorem.  
\subsection{Fano Factor}
The Fano factor $F$, or noise-to-signal ratio,  can also be of interest.  
In the case of charge shot noise, from Eq.~(\ref{electronic shot noise}) and in the low temperature limit, $F_{\rm el}:=S_{\rm el}/I_{\rm el}=e$, the charge of the charge carrier. This has been used to measure the fractional charge, e.g. $e/3,\, e/5$, of the quasiparticles predicted for a quantum Hall fluid.\cite{KanePRL94a,RdePicciottoNature97,ReznikovNature99,SaminadayarPRL97}  

Here we determine the Fano factor for a thermal current.  To simplfy things let $T_{\rm R}\to 0$, thus
\begin{equation}
F_{\rm th}:=\frac{S_{\rm th}}{I_{\rm th}}=\frac{\int_{0}^{\infty}d\epsilon\, \epsilon^{2}N^{}_{{\rm L}}(\epsilon) N^{}_{{\rm R}}(\epsilon)n^{\rm B}_{\rm L}(\epsilon)}{\int_{0}^{\infty}d\epsilon\, \epsilon\, N^{}_{{\rm L}}(\epsilon) N^{}_{{\rm R}}(\epsilon)n^{\rm B}_{\rm L}(\epsilon)}.
\end{equation}
Again assuming a power-law form of the phonon spectral density and re-scaling the integrals by letting $x=\epsilon \beta$ gives
\begin{equation}
F_{\rm th}=\frac{\int_{0}^{\infty}dx\,x^{2+2\alpha}\big[e^{x}-1\big]^{-1}}{\int_{0}^{\infty}dx\,{x^{1+2\alpha}}\big[{e^{x}-1}\big]^{-1}}k_{\rm B}T:=C(\alpha)k_{\rm B}T.
\end{equation}
Thus the Fano factor is not a universal quantity, as in the electronic case, but is  independent of all material parameters and only depends on the energy dependance of the spectral density. 
For planer surfaces\cite{PattonRBR01,GellerPRB04} $\alpha=1$ and the integrals can be done analytically giving
\begin{equation}
\label{fano factor}
F_{\rm th}=C(1)k_{\rm B}T=\frac{360\,\zeta(5)}{\pi^{4}}k^{}_{\rm B}T\approx 3.83\,k^{}_{\rm B}T,
\end{equation}
where $\zeta(x)$ is the Riemann-Zeta function.  One could loosely interpret Eq.~(\ref{fano factor}) as the average energy of the transmitted phonons through the weak link. 
\section{Discussion}
Besides the experimental ability to detect single phonons, and thus the phonon shot noise, further conditions are needed  to be in the shot noise regime. Within the model consider here, the temperature must remain well below any resonant modes of the weak link, also the link should  remain in the mesoscopic regime, i.e.~smaller than the phonon coherence length, which in itself depends on temperature. This would suggest an upper bound on temperatures of roughly $10\,{\rm K}$.  

Of course phonon noise is not only of interest for the work presented here, but could also be used to study other behavior, such as demonstrating phonon bunching in a phonon Hanbury-Brown and Twiss experiment.\footnote{M. R. Geller, private communication.}
%\begin{figure}
%\includegraphics[scale=.6]{}%
%\caption{\label{}}
%\end{figure}

\begin{acknowledgments}
The author would like to thank Michael Geller for many intriguing discussions and support from the Deutsche Forschungsgemeinschaft (DFG) under SFB 668.
\end{acknowledgments}
\bibliography{/Users/kpatton/Bibliographies/Master}
%\begin{thebibliography}{99}
%\bibitem{} 
%\end{thebibliography}

\end{document}